\documentclass[12pt]{article}

\usepackage{newtxtext,newtxmath}
\usepackage{graphicx}

\usepackage[letterpaper,margin=1in]{geometry}
\usepackage{titling}
\date{}

\makeatletter
\renewcommand{\fnum@figure}{\textbf{Figure \thefigure}}
\renewcommand{\fnum@table}{\textbf{Table \thetable}}
\makeatother
\usepackage[numbers,sort&compress,square]{natbib}
\usepackage{url}

\usepackage[colorlinks=true, linkcolor=blue, citecolor=blue, urlcolor=blue]{hyperref}
\setcitestyle{square}
\usepackage[T1]{fontenc}
\usepackage{amsmath}
\usepackage{float}
\usepackage{xcolor}
\usepackage[normalem]{ulem}
\usepackage{soul}

\newcommand{\dz}{$d_{z^{2}}$}
\newcommand{\dxy}{$d_{x^{2}-y^{2}}$}
\newcommand{\LNO}{La$_3$Ni$_2$O$_7$}

\newcommand{\LAO}{LaAlO$_3$}

\usepackage{easyReview}

\def\scititle{Spin correlations in La$_3$Ni$_2$O$_7$ thin films}
\title{\bfseries \boldmath \scititle}

\author{
	\parbox{\textwidth}{\centering
		\small
		Hengyang~Zhong$^{1\dagger}$, 
		Bo~Hao$^{2,3\dagger}$,
		Zhijia~Zhang$^{4}$,
		Anni~Chen$^{4}$,
		Yuan~Wei$^{4}$,
		Ruixian~Liu$^{1}$,
		Xinru~Huang$^{1}$,
		Chunyi~Li$^{1}$,
		Wenting~Zhang$^{1}$,
		Chang~Liu$^{1}$,
		Xiao-Sheng~Ni$^{5}$,
		Marli~dos~Reis~Cantarino$^{6}$,
		Kurt~Kummer$^{6}$,\\
		Nicholas~Brookes$^{6}$,
		Kun~Cao$^{5}$,
		Yuefeng~Nie$^{2,3,7\ast}$,
		Thorsten~Schmitt$^{4\ast}$,
		Xingye~Lu$^{1\ast}$\\[0.5em]
		\footnotesize
		$^{1}$School of Physics and Astronomy, Beijing Normal University, and Key Laboratory of Multiscale Spin Physics (Beijing Normal University), Ministry of Education, Beijing 100875, China\\
		$^{2}$National Laboratory of Solid State Microstructures, Jiangsu Key Laboratory of Artificial Functional Materials, College of Engineering and Applied Sciences, Nanjing University, Nanjing 210093, China.\\
		$^{3}$Collaborative Innovation Center of Advanced Microstructures, Nanjing University, Nanjing 210093, China.\\
		$^{4}$Photon Science Division, Swiss Light Source, Paul Scherrer Institut, CH-5232 Villigen PSI, Switzerland.\\
		$^{5}$Guangdong Provincial Key Laboratory of Magnetoelectric Physics and Devices, State Key Laboratory of Optoelectronic Materials and Technologies, Center for Neutron Science and Technology, School of Physics, Sun~Yat-Sen University, Guangzhou 510275, China.\\
		$^{6}$European Synchrotron Radiation Facility, BP 220, F-38043 Grenoble Cedex, France.\\
		$^{7}$Jiangsu Physical Science Research Center, Nanjing 210093, China.\\[0.2em]
		$^\ast$Corresponding authors. Email: ynie@nju.edu.cn; thorsten.schmitt@psi.ch; luxy@bnu.edu.cn\\
		$^\dagger$These authors contributed equally to this work.
	}
}

\begin{document}

\maketitle

\subsection*{ABSTRACT}
The discovery of ambient-pressure superconductivity with $T_{c,\text{onset}} > 40$ K in {\LNO} (LNO) thin films grown on the SrLaAlO$_4$ (SLAO) substrate with compressive ($\varepsilon\approx-2\%$) epitaxial strain provides a unique platform for investigating the superconducting mechanism in nickelate superconductors. Here, we use resonant inelastic X-ray scattering (RIXS) to unveil the dispersive spin excitations in the LNO/SLAO thin film and establish the strain dependence of the electronic and spin excitations in LNO thin films with strain ranging from $\varepsilon\approx-2\%$ to $+1.9\%$.
Compared with bulk LNO, LNO/SLAO exhibits similar $dd$ excitations and spin dynamics, but with a larger spin-excitation bandwidth,
whereas tensile-strained LNO/SrTiO$_3$ exhibits a marked suppression of both the spin excitations and the Ni $3d_{z^2}$-derived $dd$ excitations.
This evolution reflects a strain-tuned interlayer exchange interaction $J_z$ and Ni 3{\dz}-O 2$p_z$ hybridization. Our results demonstrate how epitaxial strain modulates the interlayer magnetic coupling and are consistent with scenarios in which the interlayer antiferromagnetic superexchange interaction promotes interlayer pairing in bilayer nickelates.

\noindent

\subsection*{INTRODUCTION}

The discovery of superconductivity in {\LNO} (LNO) with a superconducting onset temperature $T_{c,\rm onset} \sim 80$ K at $\sim14$ GPa hydrostatic pressure has positioned nickelates as a promising new family of high-$T_{c}$ superconductors \cite{sun2023signatures}. This breakthrough offers a unique platform for investigating the pairing mechanism of high-$T_{c}$ superconductivity \cite{ryee2024quenched, yaodx2023, liu2023s, dong2024visualization, lechermann2023electronic, sakakibara2024possible, gu2025effective, yang2023possible, lu2023superconductivity, lu2024interlayer, qu2024bilayer, oh2023type, cao2024flat, jiang2024high, yang2023interlayer, wu2024superexchange, shen2023effective, liao2023electron, zhang2024strong, yuan2024np, wang2024bulk,wang2024pressure,yao2024npj,wang2024normal,yang2024orbital}. LNO contains Ni$^{2.5+}$ ions with a $3d^{7.5}$ electron configuration \cite{sun2023signatures,mijit2024local}. The NiO$_2$ bilayers in LNO are interconnected by the inner apical oxygen (O$_{\rm AP}$) within each unit cell, leading to the presence of active Ni $3d_{z^2}$ and $3d_{x^2-y^2}$ orbitals near the Fermi level \cite{yaodx2023,yang2024orbital}, as well as significant interlayer coupling \cite{sun2023signatures}. With increasing hydrostatic pressure toward the superconducting regime, LNO undergoes a structural transition from the orthorhombic $Amam$ phase to the orthorhombic $Fmmm$ phase, while the Ni-O$_{\rm AP}$-Ni bond angle [$\varphi$ in Fig. \hyperref[fig1]{1(a)}] increases from $168^{\circ}$ to $180^{\circ}$ \cite{sun2023signatures,wang2024structure,Li2025NSR,1205_Raman}, which, together with lattice-volume compression, is proposed to modulate the hybridization between the Ni $3d_{z^2}$ and O $2p_z$ orbitals and to influence the interlayer antiferromagnetic (AFM) superexchange interaction ($J_{z}$) [Fig.~\hyperref[fig1]{1(f)}] \cite{chen2024electronic}.

Theoretically, $J_{z}$ between the adjacent NiO$_2$ layers via apical oxygen ions has been considered a key ingredient for superconductivity \cite{gu2025effective,yang2023possible,lu2023superconductivity,jiang2024high,wu2024superexchange,liao2023electron,zhang2024strong,lu2024interlayer,qu2024bilayer,oh2023type,644_Hund,cao2024flat,Chen2024,1223_Hund,Lu2025,qu2025hund,sakakibara2024possible,shen2023effective,yang2023interlayer,652_hybridization,yao2024npj,679_hybridization,1220_hybridization,1221_hybridizaiton,1222_hybridizaiton,Wang2025}. Although dispersive spin excitations of bulk {\LNO} measured by resonant inelastic X-ray scattering (RIXS) have suggested that $J_{z}$ is the dominant interaction \cite{chen2024electronic}, the evolution of spin excitations and magnetic interactions as LNO approaches its superconducting state under hydrostatic pressure remains elusive because spectroscopic measurements of spin excitations under high pressure are extremely challenging.
The recent discovery of ambient-pressure superconductivity in undoped and Pr-doped {\LNO} thin films, with $T_{c,{\rm onset}} > 40$ K in films grown on SrLaAlO$_4$ (001) (SLAO) substrates under in-plane compressive strain ($\varepsilon\sim -2\%$) \cite{ko2025signatures, zhou2025ambient,liu2025nmat,hao2025nmat,Zhou2026NSR} and $\sim$ 12 K in films grown on LaAlO$_3$ (001) (LAO) substrates under $\varepsilon\sim-1\%$ \cite{LAOsuperconductivity}, enables detailed spectroscopic investigations \cite{Sr2IrO4} of this system and provides an exciting opportunity to explore the strain-dependent evolution of the electronic structure and spin dynamics, which will help elucidate the interplay between magnetic correlations and the emergence of superconductivity in bilayer nickelates \cite{zhao2024,benjamin2024,le2025opposite,jia2025,hu2025electronic,yi2025unifying,qu2025hund}.

In this work, we use Ni-$L_3$ RIXS to explore the spin correlations in the LNO/SLAO thin film and track the evolution of electronic and spin excitations in LNO thin films epitaxially grown on SLAO, LAO, (\LAO)$_{0.3}$(Sr$_2$TaAlO$_6$)$_{0.7}$ (001) (LSAT) and SrTiO$_3$ (001) (STO) substrates. These substrates impose epitaxial strains ranging from $\varepsilon=-2\%$ to $+1.9\%$, thereby systematically modulating the overall structural framework---including the Ni-O$_{\rm AP}$-Ni bond angle $\varphi$ (from $\sim180^\circ$ to $\sim164^\circ$) and the out-of-plane lattice constant \cite{may2010, rhodes2024, zhao2024,goodge2025}. 
High-resolution RIXS measurements reveal clear dispersive spin excitations in the LNO/SLAO thin film ($\varepsilon\approx-2\%$) with $T_{c, \rm onset}\approx40$ K [Fig. \hyperref[fig1]{1(c)}], akin to bulk LNO but with a larger spin-excitation bandwidth [Fig. \hyperref[fig1]{1(d)-(e)}] and enhanced $J_{z}$ [Fig. \hyperref[fig1]{1(f)-(g)}]. This enhancement can be attributed to the optimized interlayer orbital overlap driven synergistically by in-plane compression and $c$-axis tension. In contrast, the LNO/STO film under tensile strain ($\varepsilon\approx+1.9\%$) shows a reduced spin-excitation bandwidth and strongly diminished spectral weight, together with a suppression of $dd$ excitations associated with Ni 3{\dz}-O $2p_z$ hybridization.
The well-defined, dispersive spin excitations observed in LNO/SLAO provide direct evidence for a robust interlayer antiferromagnetic exchange $J_z$, whose strain-enhanced behavior closely tracks the emergence of ambient-pressure superconductivity. 
Our results establish an empirical link between the strain-tuned structural modifications, the strength of $J_z$, and the superconducting phase. These direct spectroscopic constraints are consistent with theoretical scenarios that emphasize interlayer magnetic exchange interactions as a key ingredient for pairing, assisted by Hund's coupling \cite{lu2024interlayer,qu2024bilayer,oh2023type,644_Hund,cao2024flat,Chen2024,1223_Hund,Lu2025,qu2025hund} or \dz--\dxy\ hybridization~\cite{sakakibara2024possible,shen2023effective,yang2023interlayer,652_hybridization,yao2024npj,679_hybridization,1220_hybridization,1221_hybridizaiton,1222_hybridizaiton,Wang2025}.

\subsection*{RESULTS}

\begin{figure}
	\centering
	\includegraphics[width=0.95\textwidth]{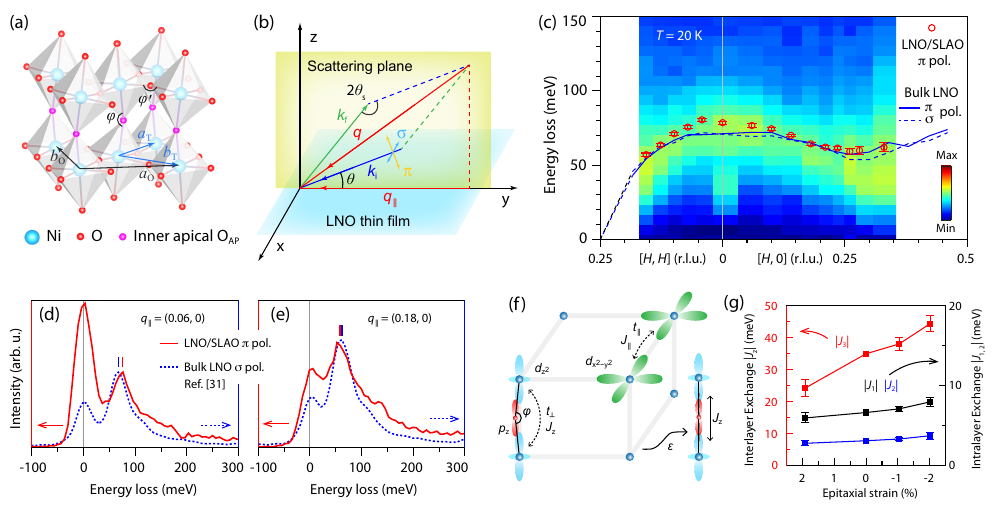}
	\caption{\textbf{Strain tuning of spin correlations in LNO thin films.} (a) Ni-O bilayer structure in LNO. The NiO$_6$ octahedra tilting causes the interlayer and intralayer Ni-O-Ni bond angles ($\varphi$ and $\varphi'$) to deviate from $180^\circ$. $a_{\rm o}$ and $b_{\rm o}$ ($a_{\rm T}$ and $b_{\rm T}$) represent the orthorhombic (pseudo-tetragonal) primitive vectors. 
(b) Scattering geometry for RIXS measurements. The scattering plane is defined by the incident vector $k_{\rm i}$ and the scattered vector $k_{\rm f}$. The electric field vector of $\pi$-polarized ($\sigma$-polarized) incident X-rays is parallel (perpendicular) to the scattering plane. $q_\parallel$ is the projection of $q$ onto the $H$-$K$ plane. 
(c) $q_\parallel$-dependent RIXS spectra of LNO/SLAO at $T=20$ K with $\pi$ polarization. Elastic scattering has been subtracted. The red circles mark the undamped energy $E_0(q_\parallel)$ for the spin excitations in LNO/SLAO, whereas the solid and dashed blue curves mark the $E_0(q_\parallel)$ fitted from $\pi$- and $\sigma$-polarized data of {\LNO} single crystal, adapted from Ref. \cite{chen2024electronic}. 
(d), (e) Comparison of spin excitations at $q_{\parallel}=(0.06, 0)$ (d) and $(0.18, 0)$ (e) between LNO/SLAO thin film (red curves) and LNO single crystal (blue dashed curves) \cite{chen2024electronic}. 
(f) Strain modulation of interlayer exchange. The synergistic variation of the Ni-O$_{\rm AP}$-Ni angle $\varphi$ and the out-of-plane distance dictates $J_z$, a key ingredient in interlayer-mediated pairing~\cite{lu2024interlayer}. Selected Ni sites show the partially filled $3d_{z^2}$/$3d_{x^2-y^2}$ orbitals and the associated hoppings ($t_\perp$, $t_\parallel$) and exchanges ($J_z$, $J_\parallel$). 
(g) Extracted values of interlayer $J_z \equiv J_3$ and intralayer $J_1$ and $J_2$ as a function of epitaxial strain. The $J_z$ value for LNO/STO represents a conservative upper bound.}
	\label{fig1}
\end{figure}

Epitaxial strain has been found to effectively tune the electronic structure and magnetic interactions in transition-metal oxides by modifying the bond lengths and bond angles \cite{1283_RIXSstrain112,Sr2IrO4}.	
We first clarify the structural framework of the LNO thin films. Bulk LNO is orthorhombic under ambient pressure, and here we adopt the pseudo-tetragonal unit cell with $a_{\rm T} = b_{\rm T} = (a_{\rm o}^{2} + b_{\rm o}^{2})^{1/2}/2$ to facilitate discussion [Fig.~\hyperref[fig1]{1(a)}]. Four substrates, SLAO, LAO, LSAT, and STO, impose nominal in-plane epitaxial strains ranging from $-2\%$ to $+1.9\%$ on LNO films, which, due to the Poisson effect \cite{1096}, result in opposite out-of-plane $c$-axis strains ranging from $+1.6\%$ to $-2.1\%$ [Fig.~S1]. Crucially, such epitaxial strain induces a synergistic structural reconstruction: it not only alters the out-of-plane Ni-Ni distances but also modulates the NiO\(_6\) octahedral tilting---a common feature of Ruddlesden-Popper nickelates that causes the Ni-O-Ni bond angle to deviate from \(180^\circ\) [Fig.~\hyperref[fig1]{1(a)}]. Previous density functional theory (DFT) calculations and scanning transmission electron microscopy (STEM) measurements consistently suggest that compressive strain drives the Ni-O-Ni bond angles closer to \(180^\circ\) while simultaneously elongating the $c$-axis, whereas tensile strain contracts the $c$-axis but enhances the octahedral tilting \cite{rhodes2024, zhao2024, yue2025, goodge2025}. This multifaceted, strain-controlled structural evolution lays the foundation for understanding the subsequent changes in electronic and spin excitations.

\begin{figure}
	\centering
	\includegraphics[width=1\textwidth]{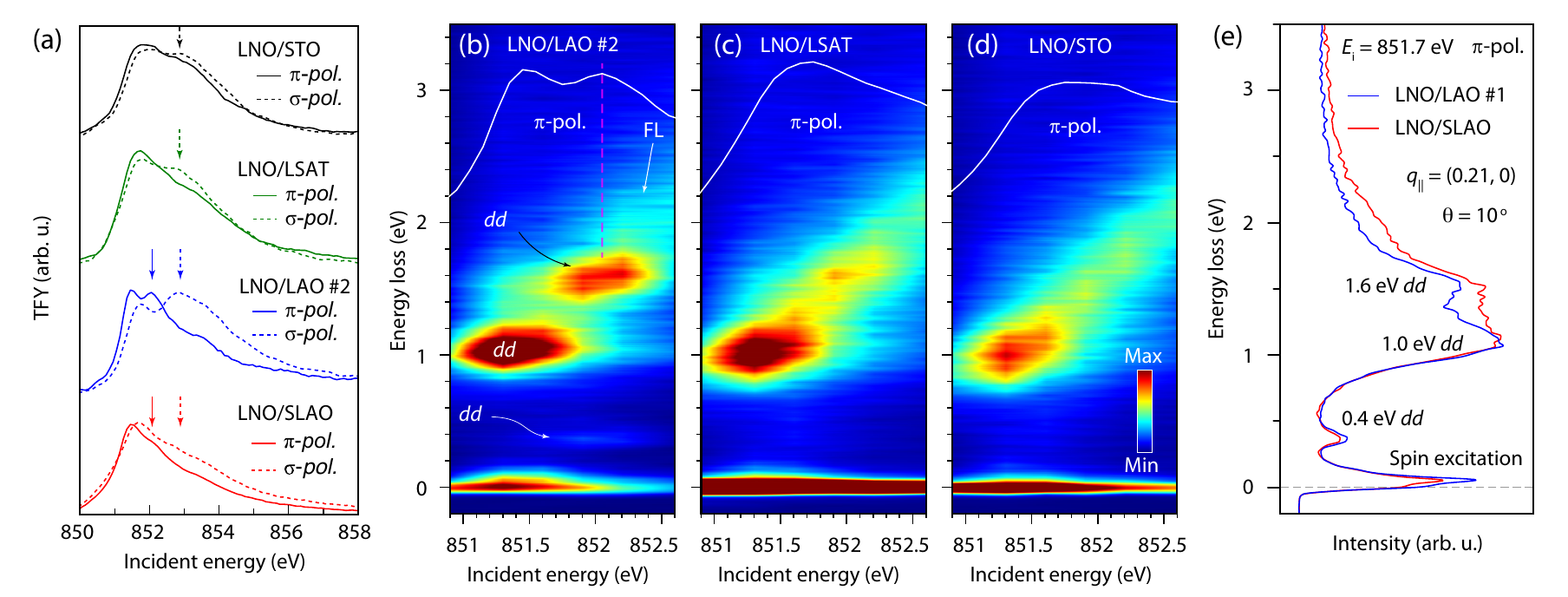}
	\caption{\textbf{Strain evolution of high-energy excitations.}  (a) XAS of LNO films near the Ni-$L_3$ edge with $\pi$ and $\sigma$ polarization after subtracting the La-$M_4$ peaks at the grazing incident angle $\theta=20^\circ$. The arrows indicate the satellite peaks, which are mainly from the $3d^8\underline{L}$ configuration \cite{chen2024electronic}. 
	(b)-(d) Incident-energy ($E_i$) dependence of the excitations of LNO films on LAO (\#2) (b), LSAT (c) and STO (d) substrates with $\pi$ polarization. The white curves are the corresponding XAS spectra and the magenta dashed line marks the resonance with the $\sim1.6$ eV excitation. All RIXS data in (b)-(d) were normalized to the incident photon flux and collected at $\theta = 20^\circ$. (e) Comparison of representative RIXS spectra of LNO/SLAO and LNO/LAO (\#1) measured at $E_{i}=851.7$ eV and $q_{\parallel}=(0.21, 0)$ with $2\theta_{s}=90^{\circ}$ and $\theta=10^\circ$.}
	\label{fig2}
\end{figure}

Figure \hyperref[fig2]{2(a)} exhibits XAS spectra of LNO thin films near the Ni-$L_3$ edge, measured at a grazing incidence angle of $\theta=20^\circ$. In this geometry, $\sigma$-polarized XAS selectively probes the unoccupied 3{\dxy} states, whereas
$\pi$-polarized XAS primarily probes the unoccupied 3{\dz} states. The $\sigma$-polarized spectra display a main resonance at $\sim851.6$ eV and a broad higher-energy satellite at $\sim852.8$ eV [dashed arrows in Fig. \hyperref[fig2]{2(a)}], which is consistent with bulk LNO \cite{chen2024electronic}. Under $\pi$ polarization, all samples show a main resonance near 851.5 eV accompanied by a high-energy satellite around 852 eV [solid arrows in Fig.~\hyperref[fig2]{2(a)}]. In both polarizations, the Ni-$L_{3}$ satellite reflects ligand-hole-rich $3d^{8}\underline{L}$ weight (O-$2p$ holes), as indicated by multiplet calculations \cite{chen2024electronic}.

Figures \hyperref[fig2]{2(b)-(d)} show the energy-dependent RIXS spectra and corresponding XAS spectra of LNO/LAO (\#2), LNO/LSAT and LNO/STO measured at $\theta=20^\circ$, respectively. In compressively strained LNO/LAO \#2 ($\varepsilon\approx-1\%$), a weak excitation located around 0.4 eV shows Raman-like characteristics. The sharp excitation located at 1 eV along with another at 1.6 eV overlaps with a broad but weak fluorescence line [Fig. \hyperref[fig2]{2(b)}]. As suggested in Ref. \cite{chen2024electronic}, the 0.4 eV and 1 eV peaks involve Ni 3{\dz} $\to$ 3{\dxy} and $e_g \to t_{2g}$ transitions [Fig.~S3], while the fluorescence-like emission stems from a delocalized Ni-O hybridized continuum. A comparison of RIXS spectra at $q_\parallel=(0.21, 0)$ [Fig. \hyperref[fig2]{2(e)}] shows that LNO/SLAO ($\varepsilon\approx-2\%$) exhibits essentially the same excitations as LNO/LAO \#1. 

The $dd$- and magnon excitations in LNO/SLAO and LNO/LAO are consistent with those in bulk LNO \cite{chen2024electronic}. As a comparison, the 0.4 eV excitation becomes significantly broadened and the 1 eV excitation displays a gradual broadening in tensile-strained LNO/LSAT ($\varepsilon\approx+0.9\%$) and LNO/STO ($\varepsilon\approx+1.9\%$) [Figs. \hyperref[fig2]{2(c), (d)}]. These broadening effects could be ascribed to the increasing metallization of the {\dz} orbital \cite{benjamin2024,zhao2024}. Specifically, the substantially reduced interlayer distance in the tensile-strained thin films drives the originally localized 3{\dz} states toward a more itinerant regime, thereby strongly damping the discrete $dd$ transitions into a broader continuum \cite{benjamin2024}.
The $\sim$1.6 eV excitation was previously suggested to stem from complex $dd$-type excitations involving Ni $3d$ and O $2p$ orbitals \cite{chen2024electronic}. Notably, this energy scale is also highly consistent with the charge-transfer gap ($\sim$1.6 eV) predicted for the interlayer 3{\dz}--O 2$p_z$ sector in La$_3$Ni$_2$O$_7$ \cite{wu2024superexchange}. Both interpretations reflect the strong interlayer hybridization mediated by the inner apical oxygen. Here, its intensity shows noticeable evolution across these samples.

As this excitation resonates at the Ni-$L_3$ edge within the $3d^8\underline{L}$ configuration (the satellite peak in XAS) [Fig. \hyperref[fig2]{2(a)}] and is strongly enhanced for $\pi$ polarization, it carries appreciable 3{\dz} weight and couples through the same 3$d^{8}\underline{L}$-rich intermediate-state channel. Thus, the gradual suppression of this excitation across LNO/SLAO, LNO/LAO, LNO/LSAT and LNO/STO can be attributed to the strain-induced overall structural modulation, which weakens the effective hybridization between the Ni 3{\dz} and inner-apical O 2$p_z$ orbitals. Concurrently, consistent with theoretical predictions \cite{benjamin2024}, the substantial $c$-axis compression under tensile strain likely drives the originally localized $3d_{z^2}$ states toward a more itinerant regime, enhancing the metallization and bandwidth \cite{benjamin2024}. Such a transition would naturally lead to strong damping of the discrete localized excitations, causing them to merge into a broader continuum—a behavior that is consistent with the suppressed peak features observed in our RIXS spectra.

\begin{figure}
	\centering
	\includegraphics[width=1\textwidth]{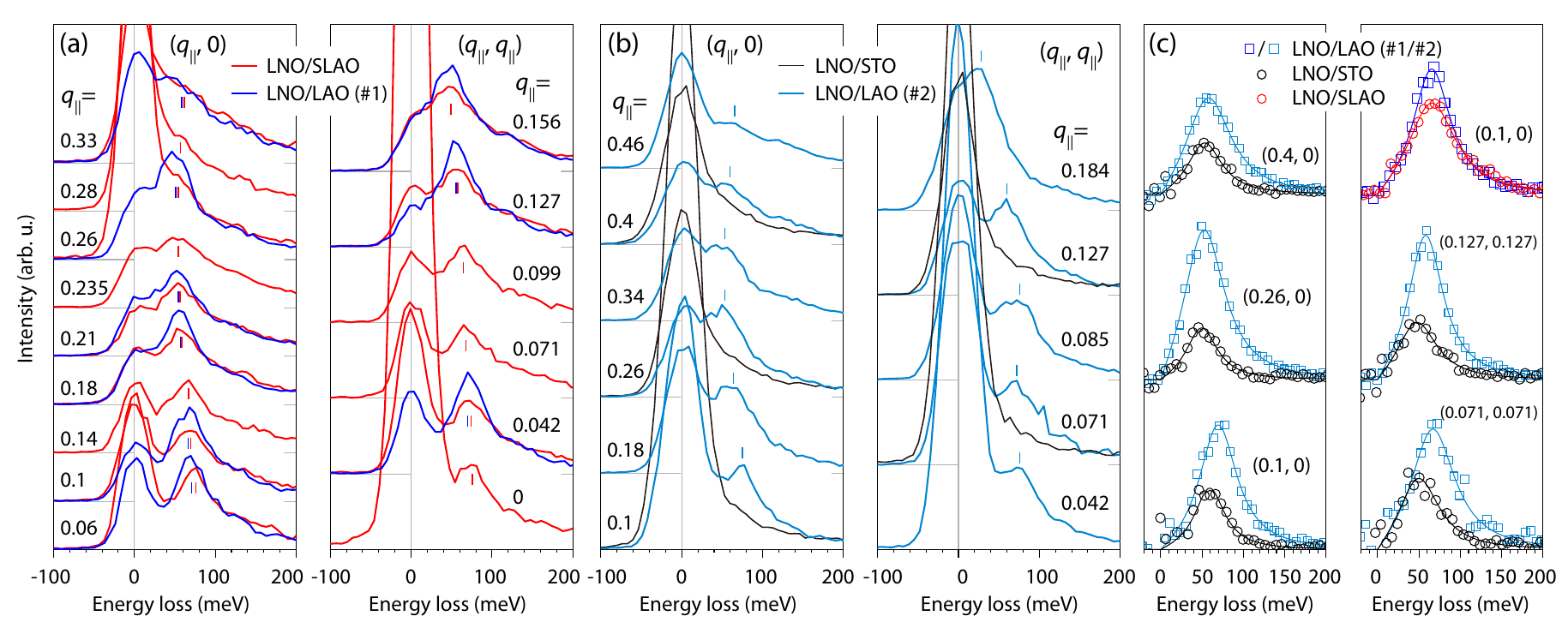}
	\caption{\textbf{Momentum-dependent spin excitations across epitaxial strain.} Comparison of $q_\parallel$-dependent RIXS spectra between LNO/SLAO and LNO/LAO (\#1) (a), and between LNO/LAO (\#2) and LNO/STO (b), measured along the $[H, 0]$ and $[H, H]$ directions with $\pi$ polarization in grazing incidence geometry. The energy resolution is $\Delta E\approx32$ meV for (a) and $\Delta E\approx45$ meV for (b). Scattering angle: in (b), $2\theta_{s}=150^\circ$; in (a), $2\theta_{s}=110^\circ$ for $|q_\parallel|\geq0.26$ and $2\theta_{s}=90^\circ$ otherwise. Vertical bars mark the peak positions of the spin excitations. (c) Spin-excitation spectra of LNO thin films extracted from the DHO fitting of the data shown in (a) and (b). The RIXS spectral intensities were normalized to the incident photon flux and unit-cell number across different samples, and to the 1 eV $dd$ excitations across different $q_{\parallel}$s.}
	\label{fig3}
\end{figure}

Having established the strain dependence of the hybridization between Ni 3{\dz} and apical O 2$p_z$ orbitals by analyzing the $dd$ excitations, we now turn to the spin excitations and their correlation with the emergence of the superconducting phase in the LNO/SLAO thin film. The $q_\parallel$-dependent RIXS spectra presented in Figs. \hyperref[fig1]{1(c)} and \hyperref[fig3]{3(a)} reveal well-defined, dispersive spin excitations along high-symmetry $[H, 0]$ and $[H, H]$ directions in LNO/SLAO at ambient pressure. The magnetic collective mode exhibits a band maximum near $\Gamma$ and disperses downward to $(1/4,1/4)$ along the $[H, H]$ direction. Its dispersion, bandwidth, and lineshape closely match those of bulk LNO [blue curves in Fig. \hyperref[fig1]{1(c)}], indicating that the underlying collinear AFM correlations \cite{Gupta2025,ren2025resolving,sdw327} are preserved in the compressively strained films. The persistence of such coherent magnetic excitations provides crucial spectroscopic constraints that are highly consistent with interlayer spin-fluctuation-mediated pairing scenarios \cite{yaodx2023,lu2024interlayer}. 
Comparable dispersive excitations are also resolved in LNO/LAO (\#1) using the same experimental setup, as shown in Fig. \hyperref[fig3]{3(a)}.

Previous studies have shown that epitaxial strain can effectively tune superconductivity in LNO thin films. Ambient-pressure superconductivity has been realized in compressively strained films grown on SLAO and LAO, with $T_{c,\mathrm{onset}} > 40$ K on SLAO \cite{ko2025signatures, zhou2025ambient,liu2025nmat,hao2025nmat,Zhou2026NSR} and $T_{c,\mathrm{onset}}\sim 12$ K on LAO \cite{LAOsuperconductivity}, whereas tensile-strained LNO/STO exhibits the lowest $T_c$ under high pressure \cite{osada2025strain}. This strain engineering thus provides an ideal platform to probe the correlation between spin excitations and the superconducting phase. 

Figure \hyperref[fig3]{3(b)} displays a comparison of RIXS spectra between LNO/LAO (\#2) and LNO/STO, measured with $\Delta E\approx45$ meV and $2\theta_{s}=150^\circ$. In LNO/LAO (\#2), clear dispersive spin excitations (cyan curves) are observed along both high-symmetry directions. By contrast, the spin excitations in LNO/STO (black curves) are strongly suppressed. To quantify the evolution of the spin excitation dispersion and spectral weight, we plot in Fig. \hyperref[fig3]{3(c)} the extracted magnon spectra fitted by a damped harmonic oscillator (DHO) function \cite{lu2022}:
$$S(q, E)=A\,\frac{4\,\gamma\,EE_0}{\left(E^2-E_0^2\right)^2+(2\gamma E\,)^2}~,$$
where $E_0(q)$ is the undamped energy and $\gamma(q)$ is the damping rate \cite{lu2022}. 

As shown in Fig. \hyperref[fig3]{3(a)}, LNO/SLAO and LNO/LAO (\#1) have comparable spin-excitation intensities throughout the measured dispersion, and a representative comparison at $q_{\parallel} = (0.1, 0)$ in Fig. \hyperref[fig3]{3(c)} shows very similar spectral weight. In sharp contrast, the overall magnetic spectral weight of LNO/STO is significantly weaker than that of LNO/LAO (\#2), indicating a collapse of magnetic correlations in the tensile-strained film. This suppression of magnetic correlations may provide a spectroscopic clue to why tensile-strained LNO/STO lies away from the robust high-$T_c$ superconducting regime in the strain-tuned phase diagram of LNO thin films \cite{osada2025strain}.

\begin{figure}
	\centering
	\includegraphics[width=0.65\textwidth]{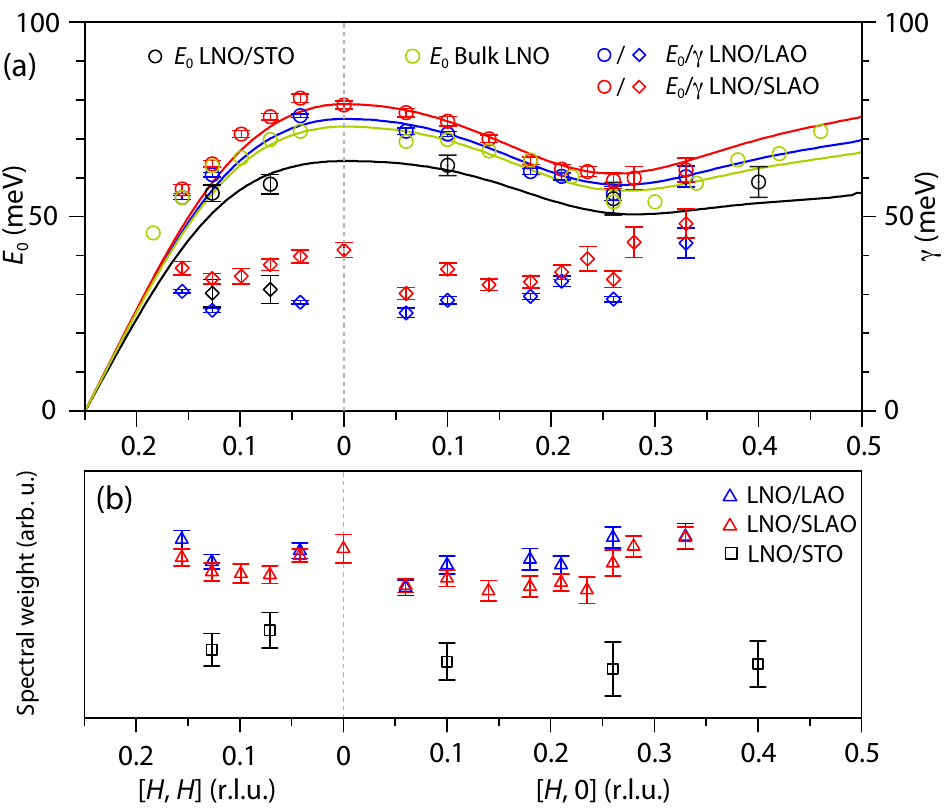}
	\caption{\textbf{Extracted magnon dispersion, damping, and spectral weight.} (a) The undamped energy $E_{0}$ (circles) of the spin excitations extracted from the fitting of the RIXS spectra using a DHO function. The damping factor $\gamma$ (diamonds) is presented only for the compressively strained LNO/SLAO and LNO/LAO (\#1) films. The red, blue, black, and green solid curves are fits to the dispersions of thin films and bulk LNO following a Heisenberg model~\cite{ni2025spin}. Additional fitting details are given in the Supplementary Materials. The bulk LNO data are taken from Ref. \cite{chen2024electronic}.
(b) Comparison of the magnon spectral weight among LNO/LAO (\#1), LNO/STO, and LNO/SLAO. The error bars in (a) and (b) were estimated from the uncertainty of the elastic peak position and the standard deviation of the fits. For clarity, only the data from the LNO/LAO (\#1) sample, measured with higher energy resolution, are shown.
	}
	\label{fig4}
\end{figure}

To obtain a quantitative understanding of the strain dependence of spin correlations, we summarize in Fig. \hyperref[fig4]{4(a)} the DHO fitting results of the spin excitations, including the undamped energy $E_{0}$ and damping factor $\gamma$ for LNO thin films, along with the spin-excitation dispersion for bulk LNO \cite{chen2024electronic}. The magnon dispersion in LNO/SLAO and LNO/LAO closely resembles that in bulk LNO, as both originate from $(0.25, 0.25)$ and reach the band top at $(0, 0)$ and $(0.5, 0)$. Despite these similarities, noticeable differences emerge near the $\Gamma$ point, where the magnon bandwidth increases by approximately 10~meV in LNO/SLAO compared with bulk LNO [Figs.~\hyperref[fig1]{1(d)} and~\hyperref[fig4]{4(a)}]. 
In stark contrast, the raw RIXS spectra of tensile-strained LNO/STO lack a distinct magnetic excitation peak. This pronounced broadening suggests a potential collapse of the coherent double-stripe spin correlations, a scenario consistent with the enhanced itinerancy and metallic behavior of this sample. Such metallization is likely driven by the substantial $c$-axis compression that pushes the 3{\dz} states closer to the Fermi level. Alternatively, a heavily suppressed magnetic mode might still exist but be largely obscured by the elastic line. To quantitatively address this ambiguity, we applied the same DHO fitting procedure to the LNO/STO spectra to establish a conservative upper bound for any surviving coherent magnetic correlations.
This upper-bound estimate indicates that the magnon energies near the $\Gamma$ point in LNO/STO are reduced by $\sim10$ meV. Furthermore, while LNO/SLAO and LNO/LAO host comparable magnon spectral weight, the extracted overall magnetic spectral weight is suppressed by $\sim70\%$ in LNO/STO [Fig.~\hyperref[fig4]{4(b)}].

Quantitatively, we use the classical Heisenberg model $H=\sum_{i<j}J_{ij}~\mathbf{S}_{i}\cdot\mathbf{S}_{j}$ to fit the energy dispersions using the SpinW code [Fig. \hyperref[fig4]{4(a)}] \cite{spinw} (See Fig.~S9 for details). The initial exchange parameters $J_1-J_{12}$ ($S_{i}$ is normalized to 1) were taken from density functional theory (DFT) calculations \cite{ni2025spin}. The interlayer coupling $J_3$ ($J_z$) and the in-plane exchanges $J_1$ and $J_2$ were treated as adjustable fitting parameters. The best fits yield a dominant $J_z = 44.4$, $38.1$, and $35.0$ meV for LNO/SLAO, LNO/LAO, and bulk LNO, respectively, alongside the conservative upper bound of $J_z \approx 24.2$ meV for LNO/STO. This indicates that $J_z$ is enhanced by approximately $27\%$ under $\varepsilon \approx-2\%$ but suppressed by at least $31\%$ under $\varepsilon \approx +1.9\%$ [Fig.~\hyperref[fig1]{1(g)}].

Theoretically, it has been suggested that $J_z$ is sensitive to changes in $\varphi$, with larger $\varphi$ enhancing $J_z$ and smaller $\varphi$ significantly suppressing it \cite{ni2025spin}. However, a recent study on tetragonal La$_3$Ni$_2$O$_{7-\delta}$ crystals featuring a bond angle of $180^\circ$ but a significantly shorter $c$-axis reported the absence of a spin-density-wave transition and quenched magnetism \cite{shi2025spin}. This implies that magnetic correlations in bilayer nickelates are sensitive to the delicate interplay between multiple structural parameters, including both the bond angle $\varphi$ and the $c$-axis lattice constant. In this context, the strain-dependent magnon dispersions and the fitted $J_z$ values [Figs. \hyperref[fig4]{4(a)} and \hyperref[fig1]{1(g)}] quantify the evolution of interlayer exchange within this complex structural landscape. This modulation is likely mediated by the effective Ni 3{\dz}-O$_{\rm AP}$ 2$p_{z}$ hybridization, which is collectively determined by the orbital alignment and the interlayer distance [Fig. \hyperref[fig2]{2}].

\subsection*{DISCUSSION}
		
Figure \hyperref[fig1]{1(f)} illustrates how epitaxial strain tunes the electronic structure. For compressively strained films, as in-plane compressive strain straightens the interlayer bond angle $\varphi$, the angular alignment optimizes the overlap between the Ni $d_{z^2}$ and O $p_z$ orbitals. This angular optimization effectively enhances the interlayer Ni--O hybridization and hopping $t_\perp$, compensating for the elongation of the out-of-plane bond length induced by the Poisson effect. This interplay accounts for the strain-dependent orbital excitations [Fig. \hyperref[fig2]{2(b)-(e)}] and interlayer exchange interactions [Fig.~\hyperref[fig1]{1(g)}]. Rather than being governed by a single parameter, this emphasizes that the electronic structure and magnetic interactions in these thin films are collectively determined by the synergistic effect of the bond angle and the interlayer distance.

Since superconductivity emerges exclusively in compressively strained films, our results are highly consistent with a spin-fluctuation-mediated pairing mechanism wherein a large interlayer exchange $J_z$ plays a vital role. In such theoretical scenarios, the enhanced $J_z$ promotes the interlayer pairing within the Ni $d_{z^2}$ orbital sector. Through Hund's rule coupling or \dz--\dxy\ hybridization, this pairing interaction is effectively transferred to the interlayer $d_{x^2-y^2}$ orbitals, which are more itinerant and weakly correlated than the \dz\ orbitals
\cite{gu2025effective,yang2023possible,lu2023superconductivity,jiang2024high,wu2024superexchange,liao2023electron,zhang2024strong,lu2024interlayer,qu2024bilayer,oh2023type,644_Hund,cao2024flat,Chen2024,1223_Hund,Lu2025,qu2025hund,sakakibara2024possible,shen2023effective,yang2023interlayer,652_hybridization,yao2024npj,679_hybridization,1220_hybridization,1221_hybridizaiton,1222_hybridizaiton,Wang2025}. Consistent with this picture, the tensile-strained LNO/STO ($\varepsilon\approx+1.9\%$) exhibits a profound suppression of the overall magnetic spectral weight and a substantially weakened $J_z$ limit. This collapse of magnetic correlations places LNO/STO far from the superconducting regime, even though it possesses a calculated Fermi surface topology similar to pressurized bulk LNO \cite{benjamin2024}.
		
It is noteworthy that the presence of a $180^\circ$ bond angle alone is insufficient for superconductivity. As recently demonstrated by Shi \textit{et al.} \cite{shi2025spin}, a fully tetragonal La$_3$Ni$_2$O$_{7-\delta}$ phase with $\varphi=180^\circ$ but a significantly shorter $c$-axis lacks both spin-density-wave order and superconductivity, highlighting that pairing requires a delicate balance of suitable electronic \cite{1266_Hall,1108_ARPES}, structural \cite{1205_Raman} conditions, and a necessary magnetic ground state. Our strain-controlled data place LNO/SLAO on the favorable side of this balance --- combining an optimized bond angle ($\varphi\approx180^\circ$) with preserved robust magnetic correlations (strong spin spectral weight and large $J_z$). Meanwhile, compared to the high-pressure bulk LNO, the LNO/SLAO film exhibits a longer $c$-axis, which likely limits the ultimate strength of $t_\perp$ and plausibly accounts for its suppressed $T_c$ relative to the bulk compound \cite{hao2025nmat}.

\subsection*{CONCLUSION}

In summary, we have utilized high-resolution Ni-$L_3$ RIXS to reveal robust, dispersive spin excitations in the LNO/SLAO thin film and establish the systematic evolution of magnetic correlations under epitaxial strain. We demonstrate that the interlayer exchange $J_z$ is not dictated by a single geometric parameter, but is instead governed by the synergistic structural modulation of the Ni-O$_{\rm AP}$-Ni bond angle $\varphi$ and the $c$-axis lattice constant, which collectively tune the effective Ni $3d_{z^2}$-O $2p_z$ hybridization. The enhancement of $J_z$ under compressive strain, and the contrasting profound collapse of overall magnetic spectral weight under tensile strain, closely track the emergence and disappearance of the superconducting phase. Ultimately, these direct spectroscopic constraints are highly consistent with pairing scenarios wherein interlayer magnetic exchange interactions, acting within a delicately balanced structural framework, serve as a key ingredient for superconductivity in bilayer nickelates.

\subsection*{METHODS}

\subsubsection*{Preparation of \LNO\ films}
LNO thin films used in this work were epitaxially grown on SLAO, LAO, LSAT, and STO substrates by reactive molecular-beam epitaxy (MBE) in a DCA R450 system. To avoid lattice relaxation, the thickness of LNO/SLAO was precisely controlled to three unit cells, whereas those of LNO/LAO, LNO/LSAT, and LNO/STO were controlled to five unit cells through layer-by-layer growth~\cite{Nie2020,hao2025nmat}. After deposition, the films were cooled to room temperature under the same oxidant pressure to suppress the formation of oxygen vacancies. LNO/SLAO was further converted into the superconducting state by a post-growth ozone annealing process. The high quality of the samples was confirmed by X-ray diffraction (XRD) and reciprocal space mapping (RSM) images [Fig.~S1].

\subsubsection*{XAS and RIXS measurements}
XAS and RIXS measurements were performed at the ID32 beamline of the European Synchrotron Radiation Facility (ESRF). All data shown were collected at $T\approx20$ K.
As shown in Fig. \hyperref[fig1]{1(b)}, the scattering plane was set to be the $yz$ plane, with $z\parallel c$ and $y$ within the $ab$ plane. XAS measurements were performed using total fluorescence yield (TFY) detection.
Incident-energy-dependent RIXS spectra were collected using $\pi$-polarized photons with an energy resolution $\Delta E\approx$ $60$ meV. Momentum-dependent RIXS spectra were collected at the Ni-$L_3$ edge using $\pi$-polarized incident photons along two high-symmetry directions $[H, 0]$ and $[H, H]$, under the grazing-incidence geometry, with $\Delta E\approx32$ meV for LNO/SLAO and LNO/LAO (\#1) and $45$ meV for LNO/LAO (\#2), LNO/LSAT, and LNO/STO.
For LNO/SLAO, the elastic scattering is significantly strong when $2\theta_s$ deviates from $90^\circ$ (see Fig. S7 for details). Therefore, most spectra were collected at $2\theta_s=90^\circ$, while $2\theta_s=110^\circ$ was used only to reach large-$q_{\parallel}$ points ($|q_{\parallel}|\geq 0.26$).

While strict precautions were taken to preserve the superconducting state of LNO/SLAO, the inevitable temporary exposure during UHV loading introduced minor oxygen deficiencies, which suppressed its macroscopic superconductivity during the RIXS measurements. Importantly, as shown in Supplementary Fig.~S8, our systematic control experiments confirm that such slight deoxygenation merely introduces damping without altering the magnon undamped energy $E_0$ or the magnetic spectral weight. Therefore, we argue that the observed spin excitations robustly reflect the intrinsic magnetic correlations of the compressively strained framework.

\subsubsection*{Spin-wave calculations}
The classical Heisenberg model \begin{equation}
	H = \sum_{i<j} J_{ij} \, \mathbf{S}_i \cdot \mathbf{S}_j
\end{equation}
was employed to fit the magnon dispersion using SpinW code, where the initial exchange parameters $J_1$-$J_{12}$ were taken from density functional theory (DFT) calculations~\cite{ni2025spin}. 
The nearest-neighbor intralayer interactions ($J_1$ and $J_2$) and the interlayer interaction ($J_3$) were treated as adjustable fitting parameters, which were refined by nonlinear least-squares fitting to the experimental magnon dispersion.
More details are provided in the Supplementary Materials.

\noindent

\clearpage

\clearpage

\clearpage

\subsection*{ACKNOWLEDGEMENTS}

We thank Yi Lu and Wei Li for helpful discussions. We acknowledge the European Synchrotron Radiation Facility (ESRF) for providing synchrotron radiation facilities under proposal numbers HC-5640 and SC-5699 at the ID32 beamline.

\subsection*{FUNDING}

The work is supported by the Scientific Research Innovation Capability Support Project for Young Faculty (ZYGXQNJSKYCXNLZCXM-M2), the National Natural Science Foundation of China (Grants no. 12574142, and 12434002), National Key Projects for Research and Development of China with Grant No. 2021YFA1400400, Natural Science Foundation of Jiangsu Province (No. BK20233001) and the National Natural Science Foundation of China (No. 125B2073), and Guangdong Basic and Applied Basic Research Foundation (Grants No. 2022A1515011618). The work at PSI is supported by the Swiss National Science Foundation through project No. 207904. Y.W. acknowledges funding from the European Union's Horizon 2020 research and innovation programme under the Marie Sklodowska-Curie grant agreement 884104 (PSI-FELLOW-II-3i program). 

\subsection*{AUTHOR CONTRIBUTIONS}

X.L., T.S., and Y.N. conceived the project. B.H. grew the thin films and carried out the transport and X-ray diffraction measurements. H.Z., Z.Z., A.C., Y.W., R.L., X.H., C.Li., W.Z., C.Liu., and X.L. performed the RIXS experiments with help from M.C., K.K., and N.B.. H.Z. and X.L. analyzed the data. K.C., X.-S.N., H.Z., and X.L. calculated and fitted the spin-excitation dispersion. H.Z. wrote the manuscript with input from X.L., B.H., Y.N., M.C., Y.W. and T.S.. All authors made comments.

\bibliographystyle{apsrev4-3}
%\bibliography{ref_LNO}
%

\end{document}